\documentclass[12pt]{spieman}  
\usepackage{amsmath,amsfonts,amssymb}
\usepackage{graphicx}
\usepackage{setspace}
\usepackage{tocloft}
\usepackage{lineno}
\usepackage{xcolor}
\usepackage{transparent}
\usepackage{multirow, multicol}
\usepackage{algorithm}
\usepackage{algpseudocode}
\usepackage{comment}
\title{Probabilistic Modeling of Photon Interaction Chains in a Silicon Detector}

\author[1,2,*]{Per Lundhammar}
\author[1.2]{Luca Terenzi}
\author[1,2]{Mats Danielsson}
\author[1,2]{Moa Yveborg Tamm}
\author[1,2]{Mats Persson}
\affil[1]{KTH Royal Institute of Technology, Department of Physics, Stockholm, Sweden}
\affil[2]{MedTechLabs, BioClinicum, Karolinska University Hospital, Solna, Sweden}

\cftpagenumbersoff{figure}
\cftpagenumbersoff{table} 
\begin{document} 
\maketitle
\begin{abstract}
Compton camera imaging relies on the positions and energies of multiple photon interactions recorded in the detector. In silicon detectors that resolve individual interactions, these measurements do not directly provide the causal interaction order, and in nuclear medicine applications the photon may also have scattered in the object before entering the detector. We present a probabilistic model for photon interaction chains in a silicon detector and use it for two inference tasks at the event level: interaction order inference and classification of primary versus object-scattered photons. The model combines photon transport based on Compton scattering and attenuation with a detector response model. It is evaluated on Monte Carlo data for monoenergetic point sources at 140~keV and 511~keV, with and without a cylindrical water phantom. For interaction chains of length two to four, the maximum a posteriori order estimate gives high accuracy in the case without a phantom, exceeding 0.95 for all chain lengths at 511~keV, and decreasing with increasing phantom radius. The gap between the two highest posterior order probabilities provides a confidence measure for the ordering estimate and can be used to increase accuracy at the cost of reduced retention. For scatter classification, likelihood-ratio testing yields receiver operating characteristic curves with area under the curve above 0.85 across all phantom radii considered. These results support the use of probabilistic modeling of measured interaction sequences for inference directly at the event level, and show that both interaction ordering and scatter discrimination can be treated within a common framework.
\end{abstract}

{\noindent \footnotesize\textbf{*}Per Lundhammar,  \linkable{lundhamm@kth.se} }

\begin{spacing}{2}   
\section{Introduction}

The Compton camera is an imaging modality similar to the traditional gamma camera, but without the mechanical collimator \cite{toddProposedCamera1974a}. It was first developed in gamma ray astronomy \cite{schonfelderTelescopeSoftGamma1973}, and has since been studied in several application areas, including environmental physics, nuclear medicine imaging, and proton therapy monitoring \cite{takedaPortableSiCdTe2015,petersonOptimizingThreestageCompton2010}. Several detector designs have also been developed, including semiconductor-based systems \cite{watanabeSiCdTeSemiconductorCompton2005,fontanaComptonCameraStudy2017,tashimaComptonImagingMedical2022}. Because the Compton camera does not use a mechanical collimator, it can detect photons from all directions, which gives a higher detection efficiency than collimated gamma cameras \cite{y.feng3DReconstructionBenchmark2020}.

A Compton camera records the positions and deposited energies of photon interactions in the detector. From Compton scattering kinematics, an interaction sequence defines a cone surface in image space on which the source must lie. By combining many such events, an estimate of the source distribution can be formed. This can be done with back projection methods and further improved with analytical, list-mode, and statistical reconstruction methods \cite{hebertThreeDimensionalMaximumLikelihood1990,sauve3DImageReconstruction1999,wildermanListmodeMaximumLikelihood1998a,wildermanListmodeEM2000,wildermanImprovedModelingSystem2001,maximProbabilisticModelsNumerical2016,tashima3DComptonImage2020}. In practice, however, the quality of the final image depends not only on the reconstruction algorithm itself, but also on how each detected event is interpreted before reconstruction.

For detectors that resolve individual interactions, the measured data do not directly specify the causal interaction order. The detector records a set of interaction positions and deposited energies, but not the order in which the interactions occurred. Since the Compton relation connects scattering angle and energy loss sequentially along the interaction chain, this missing order makes the physical interpretation of an event ambiguous. A second difficulty arises in nuclear medicine imaging, where photons may scatter in the object before entering the detector. Such object-scattered can degrade image quality if treated as primary events. These two issues, interaction ordering and object scatter, are both event-level inference problems that arise before image reconstruction and directly affect the usefulness of recorded events.

Related statistical methods for inference from incomplete or latent interaction information have been developed in nuclear imaging, where unknown event attributes are inferred from measured detector signals or event data \cite{champleyStatisticalLOREstimation2009,idaAccurateModelingEventbyevent2019}. For Compton systems, previous work has studied tasks such as interaction sequence reconstruction, event selection, and improved physical modeling of the measurement process \cite{leeEvaluationSequenceTracking2021,draegerFeasibilityStudiesNew2017,fengInfluenceDopplerBroadening2021,zoglauerDopplerBroadening2003,yonedaReconstructionMultipleCompton2023}. Recent work has also explored machine-learning-based approaches for event reconstruction and event selection \cite{takashimaEventReconstruction2022,perezCurbeloImprovingCompton2025,danielDeepLearningCompton2022}. However, there remains a need for event-level models that are explicitly physics-based, probabilistic, and flexible enough to support multiple inference tasks within a common framework.

In this work, we present a probabilistic model for photon interaction chains in a silicon detector. The model combines photon transport physics with a detector response model to assign likelihood values to candidate interaction chains. While related probabilistic ideas have appeared in earlier work, the present formulation is tailored to event-level inference in a silicon detector and is used here in two ways. First, to infer the most likely causal interaction order from an unordered set of measured interactions. Secondly, to distinguish between primary and object-scattered photons. The main contribution of this work is therefore a unified probabilistic framework for interpreting recorded interaction sets before image reconstruction. In this sense, the work is not primarily a reconstruction method, but a physics-based event-level inference framework that can support reconstruction, event selection, and related downstream processing.
\section{Methods}

\subsection{Model overview}
\begin{figure}[h]
    \centering
    \includegraphics[width=1\linewidth]{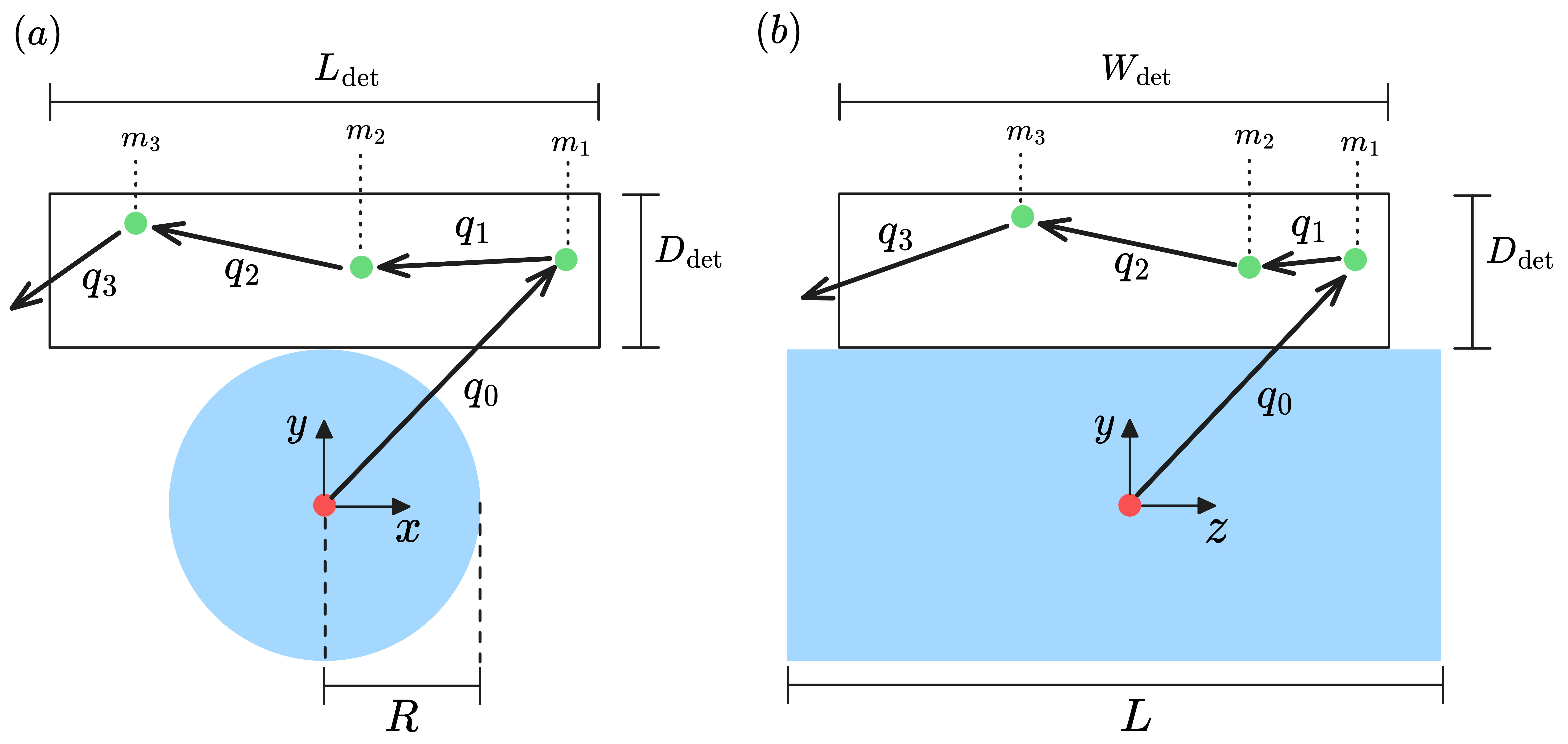}
    \caption{
Schematic overview of the detector--phantom configuration and photon interaction model.
(a) $xy$-plane and (b) $zy$-plane. The detector has dimensions $L_{\mathrm{det}}$, $W_{\mathrm{det}}$, and $D_{\mathrm{det}}$, and a global Cartesian coordinate system is indicated. A photon enters the detector with state $q_0$ and undergoes a sequence of interactions, producing a latent interaction chain $q_0 \rightarrow q_1 \rightarrow q_2 \rightarrow q_3$. The detector records an unordered set of measurements $m = \{m_1, m_2, m_3\}$. Each measurement corresponds to one interaction, but the detector does not encode the causal order of those interactions.
}
    \label{fig:overview}
\end{figure}

We consider a silicon detector defined in a global Cartesian coordinate system. The detector records the positions and deposited energies of photon interactions occurring within the detector volume. A photon enters the detector at position $\mathbf{r}_0 \in \mathbb{R}^3$ with direction $\mathbf{u}_0 \in S^2$ and energy $E_0 \in \mathbb{R}_+$. The photon state at detector entry is denoted by $q_0 = (\mathbf{r}_0, E_0, \mathbf{u}_0)$. Assume that the photon undergoes $n \geq 1$ interactions inside the detector. After interaction $i$, the latent photon state is denoted by $q_i = (\mathbf{r}_i, E_i, \mathbf{u}_i)$ and consists of the interaction position $\mathbf{r}_i$, the photon energy $E_i$, and the outgoing direction $\mathbf{u}_i$. The sequence of latent states forms an interaction chain, denoted by $q = (q_1, \dots, q_n)$.

At interaction $i$, the photon deposits energy $\varepsilon_i$. For interactions $i=1,\dots,n-1$, this is given by $\varepsilon_i = E_{i-1} - E_i$. The final deposited energy $\varepsilon_n$ is determined by the terminal transition model. For a transition from state $q_{i-1}$ to $q_i$, the detector records a measurement $m_i = (\hat{\mathbf{r}}_i, \hat{\varepsilon}_i)$, where $\hat{\mathbf{r}}_i$ and $\hat{\varepsilon}_i$ are the measured interaction position and measured deposited energy, corresponding to $\mathbf{r}_i$ and $\varepsilon_i$, respectively; see Fig.~\ref{fig:overview}. The detector does not provide the causal order of the interactions. Instead, the recorded data is an unordered collection of measurements, denoted by $m = \{m_1, \dots, m_n\}$, where the indices serve only as labels and do not represent the order in which the interactions occurred. To represent a candidate ordering of the measurements, we introduce a permutation $\tau$. The ordered measurement sequence induced by $\tau$ is denoted by $m_\tau = (m_{\tau(1)}, \dots, m_{\tau(n)})$.

\subsection{Photon transport model}

Consider a chain of interactions $q = (q_1, \dots, q_n)$, and specifically the transition from state $q_{i-1}$ to $q_i$ with $i < n$. The conditional probability density of the scattered photon energy $E_i$ given incoming energy $E_{i-1}$ is derived from the Klein--Nishina differential cross section and can be written as
\begin{equation}
\begin{aligned}
p(E_i \mid E_{i-1})
&= \mathbf{1}(E_i,E_{i-1})
\frac{C(E_{i-1})}{E_i E_{i-1}}
\Bigg[
1 + \left(\frac{E_i}{E_{i-1}}\right)^2 \\
&\quad +
\frac{E_i}{E_{i-1}}
\left(
\left(
\frac{mc^2}{E_{i-1}} - \frac{mc^2}{E_i} - 1
\right)^2 - 1
\right)
\Bigg].
\end{aligned}
\label{eq:K-N_energy}
\end{equation}where $C(E_{i-1})$ ensures normalization \cite{klein_nishina}. Here $\mathbf{1}(E_i,E_{i-1})$ is equal to one if $mc^2 E_{i-1}/(mc^2+2E_{i-1}) \leq E_i < E_{i-1}$ and zero otherwise. This ensures that the scattered energy satisfies the Compton equation, $E_i = mc^2 E_{i-1}/(mc^2 + E_{i-1}(1-\cos\theta_i))$, where $\cos\theta_i = \langle \mathbf{u}_{i-1}, \mathbf{u}_i \rangle$ and $\langle \cdot,\cdot\rangle$ denotes the scalar product \cite{Compton_equation}.

Let $\mu_c(E)$ and $\mu_p(E)$ denote the Compton and photoelectric attenuation coefficients, and let $\mu(E) = \mu_c(E) + \mu_p(E)$ denote the total attenuation coefficient. Conditional on the photon propagation direction, the distance to the next interaction is modeled by Beer--Lambert attenuation. Thus, for a transition from $\mathbf{r}_{i-1}$ to $\mathbf{r}_i$, we write
\begin{equation}
\begin{aligned}
  p(\mathbf{r}_i \mid \mathbf{r}_{i-1}, E_{i-1})
&=
C'(\mathbf{r}_{i-1}, E_{i-1})\mu_c(E_{i-1})\\
&\times e^{-\mu(E_{i-1})\|\mathbf{r}_i-\mathbf{r}_{i-1}\|},  
\end{aligned}
\end{equation}
where $C'(\mathbf{r}_{i-1}, E_{i-1})$ ensures normalization over the admissible detector geometry. Thus, this factor should be interpreted as the probability density for the distance to the next interaction along the photon ray, rather than as an isotropic density over all points in space.

Combining these two effects, we write the transition probability for $i<n$ as $p(q_i \mid q_{i-1}) = p(E_i \mid E_{i-1})\, p(\mathbf{r}_i \mid \mathbf{r}_{i-1}, E_{i-1})$. For the last state transition we write
\begin{equation}
\begin{aligned}
p(q_n \mid q_{n-1})
&= C''(q_{n-1})
\Big[
\mu_p(E_{n-1})\delta(E_n) \\
&\qquad +
\mu_c(E_{n-1})p(E_n \mid E_{n-1})
\Big]
e^{-\mu(E_{n-1})\|\mathbf{r}_n-\mathbf{r}_{n-1}\|}.
\end{aligned}
\end{equation}where $\delta(\cdot)$ is the Dirac delta function and $C''(q_{n-1})$ ensures normalization, taking into account the possibility of photoelectric absorption in the last interaction. The conditional probability of an interaction chain $q$ given an initial state $q_0$ is then written as
$p(q \mid q_0) = \prod_{i=1}^{n} p(q_i \mid q_{i-1})$.
\subsection{Measurement model}

The measurement model is described by the conditional probability of measuring the chain $m$ given the corresponding latent chain $q$, denoted by $p(m \mid q)$. In the present work, we adopt an idealized detector model in order to isolate the event-level information content of the interaction sequence before introducing additional detector non-idealities. The assumptions are as follows:

\begin{enumerate}
\item \textbf{Perfect position measurement.}  
The interaction position is assumed to be measured without error, that is, $p(\hat{\mathbf{r}}_i \mid \mathbf{r}_i) = \delta(\hat{\mathbf{r}}_i - \mathbf{r}_i)$.

\item \textbf{Energy uncertainty due to Doppler broadening.}  
For interactions $i=1,\dots,n-1$, energy degradation is modeled as Gaussian broadening due to the Doppler motion of bound electrons in silicon:
\begin{equation}
p(\hat{\varepsilon}_i \mid \varepsilon_i,\cos\theta_i)
=
\frac{1}{\sqrt{2\pi}\,\sigma_{\mathrm{db}}}
\exp\left(
-\frac{(\hat{\varepsilon}_i - \varepsilon_i)^2}{2\sigma_{\mathrm{db}}^2}
\right),
\end{equation}
where $\sigma_{\mathrm{db}} = \sigma_{\mathrm{db}}(\varepsilon_i,\cos\theta_i)$. The function $\sigma_{\mathrm{db}}$ is obtained by fitting a polynomial model to Monte Carlo simulations of tabulated values of the standard deviation in deposited energy due to Doppler broadening over different scattering angles and incident energies \cite{Compton_coincidence}. For the final interaction, the measured deposited energy is modeled by a terminal likelihood $p(\hat{\varepsilon}_n \mid \varepsilon_n)$ that depends only on the deposited energy.

\item \textbf{Conditional independence.}  
Measurement errors are assumed conditionally independent between interactions, so that
\begin{equation}
p(m \mid q)
=
\prod_{i=1}^{n} p(m_i \mid q_{i-1}, q_i).
\end{equation}
For $i=1,\dots,n-1$, we write
\begin{equation}
p(m_i \mid q_{i-1}, q_i)
=
p(\hat{\varepsilon}_i \mid \varepsilon_i,\cos\theta_i)\, p(\hat{\mathbf{r}}_i \mid \mathbf{r}_i),
\end{equation}
whereas for the final interaction,
\begin{equation}
p(m_n \mid q_{n-1}, q_n)
=
p(\hat{\varepsilon}_n \mid \varepsilon_n)\, p(\hat{\mathbf{r}}_n \mid \mathbf{r}_n).
\end{equation}
\end{enumerate}

Other detector effects such as finite pixel resolution, charge sharing, and electronic noise are neglected.

\subsection{Marginal likelihood}
\label{sec:marignal_likelihood}

The likelihood of the measured interaction chain given the initial photon state is obtained by marginalizing over latent interaction chains,
\begin{equation}
p(m \mid q_0)
=
\int p(q \mid q_0)\, p(m \mid q)\, dq,
\end{equation}
with $dq = \prod_{i=1}^n dq_i = \prod_{i=1}^n d\mathbf{r}_i\, dE_i\, d\mathbf{u}_i$.

Under the assumption of perfect position measurement, the latent interaction positions coincide with the measured ones, that is, $\mathbf{r}_i = \hat{\mathbf{r}}_i$ for $i=1,\dots,n$. For a fixed ordered measurement chain $m_\tau$, the ordered interaction geometry is therefore determined by the measured positions. In particular, the propagation directions between consecutive interactions satisfy $\mathbf{u}_{i-1} = (\mathbf{r}_i-\mathbf{r}_{i-1})/\|\mathbf{r}_i-\mathbf{r}_{i-1}\|$ for $i=1,\dots,n$, and hence the scattering angles are determined by the ordered geometry through $\cos\theta_i = \langle \mathbf{u}_{i-1}, \mathbf{u}_i \rangle$.
Given the incident energy $E_0$, the post-interaction photon energies are then determined recursively by the Compton equation,
\begin{equation}
E_i
=
\frac{mc^2 E_{i-1}}{mc^2 + E_{i-1}(1-\cos\theta_i)},
\qquad i=1,\dots,n-1.
\end{equation}
For interactions $i=1,\dots,n-1$, the deposited energies satisfy $\varepsilon_i = E_{i-1}-E_i$ and are therefore also determined recursively by $E_0$ and the ordered measured geometry. The final deposited energy $\varepsilon_n$ is determined by the terminal transition model. Thus, for a fixed ordered measurement chain $m_\tau$, there exist deterministic functions $E_i = E_i(E_0,m_\tau)$ for $i=1,\dots,n-1$ and $\varepsilon_i = \varepsilon_i(E_0,m_\tau)$ for $i=1,\dots,n$. 

As a consequence, once the ordered interaction positions are fixed by the measurements, the latent interaction chain is completely determined by the detector-entry state. Let $q(m_\tau,q_0)$ denote the unique latent interaction chain compatible with the ordered measurements $m_\tau$ and the entry state $q_0$. Then $p(m_\tau \mid q_0) = p(q(m_\tau,q_0)\mid q_0)\, p(m_\tau \mid q(m_\tau,q_0))$.

Furthermore, since $\mathbf{r}_1$ is fixed by the measurement, the detector-entry point $\mathbf{r}_0$ is determined by the incoming direction $\mathbf{u}_0$ together with the detector geometry. The remaining continuous degrees of freedom in the likelihood are therefore the incident energy $E_0$ and the incident direction. By introducing a local coordinate system centered at the first interaction position, the likelihood may equivalently be parameterized by $(\mathbf{u}_0', E_0)$, where $\mathbf{u}_0' = -\mathbf{u}_0$ denotes the outgoing direction from the first interaction position $\mathbf{r}_1$. See Appendix A for the explicit expression of $p(m_\tau \mid E_0,\mathbf{u}_0')$ under this parametrization and measurement model.

\subsection{Priors}
\label{sec:priors}

Both the order inference and scatter inference tasks require prior distributions over the photon state at detector entry. The prior over permutations is taken to be uniform, that is, $p(\tau) \propto 1$.

The prior over the incident direction is defined through a geometric source model. We parameterize the direction using $\mathbf{u}_0'$, representing the outgoing direction in a local coordinate system centered at the first interaction position $\mathbf{r}_1$, as in Section~\ref{sec:marignal_likelihood}. Let $\Omega \subset \mathbb{R}^3$ denote a bounding region in which the source distribution is embedded. For a given direction $\mathbf{u}_0'$, consider the ray originating at $\mathbf{r}_1$ and extending in direction $\mathbf{u}_0'$. The directional prior is defined as
\begin{equation}
p(\mathbf{u}_0') \propto \int_0^\infty \mathbf{1}_{\Omega}(\mathbf{r}_1 + t \mathbf{u}_0') \, dt,
\end{equation}
where $\mathbf{1}_{\Omega}(\mathbf{r}_1+t\mathbf{u}_0')$ equals one if $\mathbf{r}_1+t\mathbf{u}_0' \in \Omega$ and zero otherwise. Thus, the prior is proportional to the length of the intersection between the back-projected ray and the source region $\Omega$. The resulting density is normalized over the unit sphere. This construction corresponds to assuming a uniform spatial source distribution within $\Omega$.

In the presence of a water phantom, the source region $\Omega$ is taken to be a finite cylinder with radius $R$ and length $L$, aligned with the phantom geometry. In the absence of a phantom, corresponding to $R=0$, the directional prior is taken to be isotropic, $p(\mathbf{u}_0') = 1/(4\pi)$.

When photons traverse an object before entering the detector, the photon energy at detector entry depends on the number of Compton scatter events occurring in the object. Let $p_1(E' \mid E)$ denote the single-scatter energy distribution in Eq.~\eqref{eq:K-N_energy}, that is, the conditional density of the outgoing photon energy $E'$ after one Compton scatter given incident energy $E$. The energy distribution after exactly $n$ object scatter events is defined recursively as
\begin{equation}
p_n(E \mid E_0)
=
\int p_1(E \mid E')\, p_{n-1}(E' \mid E_0)\, dE',
\end{equation}
with $p_0(E \mid E_0) = \delta(E-E_0)$.

To obtain the prior over the photon energy at detector entry, these conditional energy distributions are weighted by the probability that a photon exits the object after exactly $n$ scatter events. Let $w_n$ denote this probability. The resulting energy prior is modeled as
\begin{equation}
p_{\mathrm{exit}}(E \mid E_0)
=
\sum_{n=0}^{\infty} w_n\, p_n(E \mid E_0),
\end{equation}
where $\sum_{n=0}^{\infty} w_n = 1$. The weights $w_n$ are not known a priori and depend on the object geometry, material composition, and source configuration. In the present work, they are therefore treated as geometry-dependent quantities and estimated using Monte Carlo simulations for the phantom under consideration. In practice, the infinite series is truncated at a finite number of scatter orders, and the corresponding truncated weights are renormalized. The estimated scatter-order fractions used in the prior construction are reported together with the simulation results.

\subsection{Posterior}

Given a measured interaction set $m$, inference is formulated in terms of the posterior distribution over the unknown interaction order $\tau$ and the photon state at detector entry. We assume that the priors over interaction order, incident energy, and incident direction are independent, so that $p(\tau,E_0,\mathbf{u}_0') = p(\tau)\,p(E_0)\,p(\mathbf{u}_0')$. Using Bayes' rule, the posterior can then be written as
\begin{equation}
p(\tau, E_0, \mathbf{u}_0' \mid m)
\propto
p(m_\tau \mid E_0, \mathbf{u}_0')\,
p(E_0)\,
p(\mathbf{u}_0')\,
p(\tau),
\end{equation}
where $p(m_\tau \mid E_0, \mathbf{u}_0')$ is the likelihood of the ordered measurement sequence induced by $\tau$ and parameterized by $(E_0, \mathbf{u}_0')$.

The interaction order is treated as a latent discrete variable. The posterior over $\tau$ is obtained by marginalizing over the continuous parameters,
\begin{equation}
p(\tau \mid m)
\propto
p(\tau)
\int
p(m_\tau \mid E_0, \mathbf{u}_0')\,
p(E_0)\,
p(\mathbf{u}_0')\,
dE_0\, d\mathbf{u}_0'.
\end{equation}
The most probable ordering is obtained using a maximum a posteriori (MAP) estimate,
\begin{equation}
\tau^*
=
\arg\max_{\tau}
p(\tau \mid m).
\end{equation}
The accuracy is defined as the fraction of correctly inferred interaction orders.

The posterior distribution over interaction orders provides a measure of inference confidence. Let $\tau_1 = \arg\max_{\tau} p(\tau \mid m)$ denote the MAP interaction order, and let $\tau_2 = \arg\max_{\tau \neq \tau_1} p(\tau \mid m)$ denote the second most probable order. We define the log-probability gap as
\begin{equation}
\Delta \log p
=
\log p(\tau_1 \mid m) - \log p(\tau_2 \mid m).
\end{equation}
This quantity corresponds to the logarithm of the posterior odds ratio between the two most probable ordering hypotheses and provides a measure of how strongly the data support the MAP solution relative to its closest competitor. The log-probability gap can be used to select events with high inference confidence. Given a threshold $T$, events are retained only if $\Delta \log p \geq T$. Increasing the threshold improves the accuracy among retained events, at the cost of reducing the fraction of accepted events.

Considering scattered photons, we say that a photon that has not scattered in the object before entering the detector is a primary photon. To distinguish between primary photons and non-primary photons, that is, photons that have scattered at least once before entering the detector, we consider two competing hypotheses on the measured set $m$:
\begin{equation*}
    \begin{aligned}
        H_0:&~m~\text{is the measurement of a primary photon},\\
        H_1:&~m~\text{is the measurement of a non-primary photon.}
    \end{aligned}
\end{equation*}
The hypotheses differ only in the prior over the incident energy $E_0$. Under $H_0$, corresponding to primary photons, the incident energy prior is modeled as a monoenergetic distribution, $p(E_0 \mid H_0) = \delta(E_0 - E_{\mathrm{src}})$, where $E_{\mathrm{src}}$ is the known source energy. Under $H_1$, corresponding to non-primary photons, the incident energy prior is given by the mixture distribution defined in Section~\ref{sec:priors}, conditioned on one or more scatter events,
\begin{equation}
p(E_0 \mid H_1)
=
\frac{\sum_{n=1}^{N} w_n\, p_n(E_0 \mid E_{\mathrm{src}})}{\sum_{n=1}^{N} w_n}.
\end{equation}
The likelihood of each hypothesis is obtained by marginalizing over $(\tau, E_0, \mathbf{u}_0')$,
\begin{equation}
\begin{aligned}
p(H_k \mid m)
&\propto
\sum_{\tau}
\int
p(m_\tau \mid E_0,\mathbf{u}_0')\,
p(E_0 \mid H_k)\,
p(\mathbf{u}_0')\,
p(\tau) \\
&\qquad \times dE_0\, d\mathbf{u}_0',
\qquad k \in \{0,1\}.
\end{aligned}
\end{equation}
To perform scatter classification, we define the likelihood ratio in favor of the primary hypothesis $H_0$ relative to the non-primary hypothesis $H_1$ as
\begin{equation}
\Lambda(m)
=
\frac{p(m \mid H_0)p(H_0)}
     {p(m \mid H_1)p(H_1)}.
\end{equation}
where $p(H_i)$ is the prior probability of hypothesis $i$. Classification is performed by thresholding the likelihood ratio, that is, we label $m$ as a primary photon if $\Lambda(m) > \eta$, where $\eta$ is a decision threshold. The performance of the classifier is evaluated using receiver operating characteristic (ROC) curves, obtained by varying the threshold $\eta$ and computing the corresponding true positive and false positive rates. The overall discriminative performance is summarized by the area under the curve (AUC).

In the experiments presented in this work, scatter inference is performed under the assumption that the interaction order is known. That is, the likelihood is evaluated for a fixed ordering $\tau$, and no marginalization over permutations is performed. This corresponds to a cascaded approach in which interaction ordering is performed as a preprocessing step.

\subsection{Simulation setup}

\paragraph{General setup}
The detector is modeled as a silicon volume with dimensions $L_{\mathrm{det}} = W_{\mathrm{det}} = 20\,\mathrm{cm}$ and depth $D_{\mathrm{det}} = 5\,\mathrm{cm}$. A cylindrical water phantom of length $L = 30\,\mathrm{cm}$ and radius $R \in \{0,5,10,15\}\,\mathrm{cm}$ is placed directly in front of the detector, with the phantom surface touching the incident detector plane for $R>0$, see Fig.~\ref{fig:overview}. The case $R=0$ corresponds to the absence of a phantom; in that case, the point source is placed at a perpendicular distance of $15\,\mathrm{cm}$ from the incident detector plane. A monoenergetic point source emits photons with energy $E_{\mathrm{src}} \in \{140,511\}\,\mathrm{keV}$. All Monte Carlo datasets were generated using \textsc{Allpix}$^2$~\cite{spannagelAllpix2ModularSimulation2018}.

\paragraph{Order inference}
For the order inference task, $3000$ events were used for each configuration of source energy and phantom radius, balanced across interaction chain lengths $n \in \{2,3,4\}$ with $1000$ events per chain length. Thus, for a given phantom configuration, the proportion of primary and non-primary photons in the dataset reflects the proportions induced by that configuration in the underlying Monte Carlo simulation; no balancing was performed with respect to phantom scatter status. For each configuration, the interaction order is inferred as described above, and the accuracy is computed as the fraction of correctly inferred orders. To assess confidence-based selection, a threshold is applied to the log-probability gap $\Delta \log p$. The retention is defined as the fraction of events satisfying $\Delta \log p \geq T$, and the accuracy is evaluated on the retained subset as a function of the threshold $T$.

\paragraph{Scatter classification}
For the scatter classification task, $1000$ events were used for each configuration with $R>0$, balanced across the two classes primary and non-primary, with $500$ events in each class. These events were selected from the larger Monte Carlo dataset. The non-primary class therefore inherits the energy distribution produced by photon transport through the phantom in the underlying Monte Carlo simulation. Thus, unlike the order-inference datasets, the scatter-classification datasets were deliberately balanced with respect to scatter status, while remaining unbalanced with respect to interaction chain length. This balanced design was chosen because the natural prevalence of non-primary photons depends strongly on phantom size. In an extreme case with a very small phantom, a trivial classifier that labels nearly every event as primary could appear favorable simply because most events are primary, which would obscure the actual event-level scatter differentiation problem. By balancing the two classes, the evaluation instead focuses on how well the measured interaction sequence separates primary from non-primary photons. Scatter classification is performed using the likelihood ratio test described above, and performance is evaluated using receiver operating characteristic curves and the corresponding area under the curve.
\section{Results}

\subsection{Evaluation datasets}

The results are reported separately for the two evaluation tasks. For the order-inference task, validation datasets were constructed with 3000 events for each configuration of source energy and phantom radius, balanced across interaction-chain lengths $n \in \{2,3,4\}$ with 1000 events per chain length. Table~\ref{tab:non_primary_fraction} reports, for the phantom configurations, the fraction of non-primary photons within each chain-length subset. Thus, the reported percentages are computed from the selected order-inference validation sets and reflect the proportion of photons that underwent scatter in the phantom within each subset.

For the scatter-classification task, validation datasets were constructed with 1000 events for each configuration with $R>0$, balanced across the two classes primary and non-primary, with 500 events in each class. Table~\ref{tab:scatter_chain_lengths} summarizes the distribution of interaction-chain lengths in these balanced datasets. Since the datasets were constructed to contain equal numbers of primary and non-primary photons, the table reflects the chain-length composition of the selected evaluation sets rather than the natural class prevalence in the underlying Monte Carlo simulations.

\begin{table}[t]
\centering
\caption{Fraction of non-primary photons in the order-inference validation datasets for interaction-chain lengths $n=2$, $n=3$, and $n=4$. For each combination of source energy and phantom radius, the validation dataset contains 1000 events for each chain length.}
\label{tab:non_primary_fraction}
\begin{tabular}{ccccc}
\hline
$E_{\mathrm{src}}$ (keV) & $R$ (cm) & $n=2$ (\%) & $n=3$ (\%) & $n=4$ (\%) \\
\hline
\multirow{3}{*}{140}
& 5  & 68.9 & 63.6 & 59.7 \\
& 10 & 86.8 & 84.8 & 78.6 \\
& 15 & 92.0 & 88.0 & 87.7 \\
\hline
\multirow{3}{*}{511}
& 5  & 56.0 & 61.1 & 60.4 \\
& 10 & 79.9 & 78.7 & 79.2 \\
& 15 & 86.5 & 89.1 & 88.7 \\
\hline
\end{tabular}
\end{table}
\begin{table*}[!t]
\centering
\caption{Distribution of interaction-chain lengths in the balanced validation datasets used for scatter classification. For each combination of source energy and phantom radius, the dataset contains 1000 events in total, with 500 primary and 500 non-primary photons.}
\label{tab:scatter_chain_lengths}
\begin{tabular}{ccccccc}
\hline
$E_{\mathrm{src}}$ (keV) & $R$ (cm) & $n=1$ (\%) & $n=2$ (\%) & $n=3$ (\%) & $n=4$ (\%) & $n>4$ (\%) \\
\hline
\multirow{3}{*}{140}
& 5  & 36.6 & 24.8 & 15.4 & 10.7 & 12.5 \\
& 10 & 39.3 & 24.0 & 14.9 & 8.5  & 13.3 \\
& 15 & 38.9 & 24.8 & 15.9 & 10.2 & 10.2 \\
\hline
\multirow{3}{*}{511}
& 5  & 39.3 & 24.1 & 12.8 & 9.3  & 14.5 \\
& 10 & 45.4 & 20.6 & 12.5 & 7.3  & 14.2 \\
& 15 & 47.0 & 20.3 & 12.7 & 6.8  & 13.2 \\
\hline
\end{tabular}
\end{table*}

\subsection{Interaction order inference}

\begin{figure}[ht!]
\centering
\includegraphics[width=\linewidth]{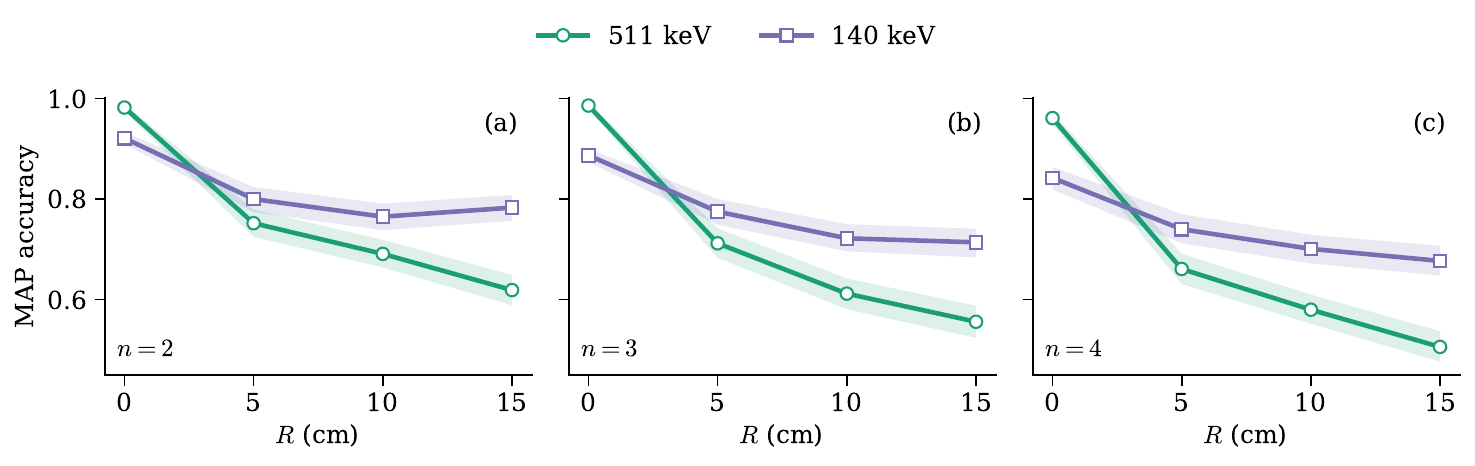}
\caption{Baseline MAP interaction-order inference accuracy as a function of phantom radius for interaction chains of length $n=2$, $n=3$, and $n=4$. No confidence filtering is applied. Shaded bands indicate bootstrap 95\% confidence intervals, obtained by resampling events within each evaluation dataset.}
\label{fig:order_accuracy_vs_radius}
\end{figure}

Figure~\ref{fig:order_accuracy_vs_radius} shows the baseline MAP interaction-order inference accuracy as a function of phantom radius for interaction-chain lengths $n=2$, $n=3$, and $n=4$, and for both source energies. No confidence filtering is applied in this figure. For all interaction lengths, the ordering accuracy decreases with increasing phantom radius. For a fixed phantom radius, the accuracy also decreases with increasing interaction-chain length.

\subsection{Confidence filtering}

\begin{figure}[ht!]
\centering
\includegraphics[width=\linewidth]{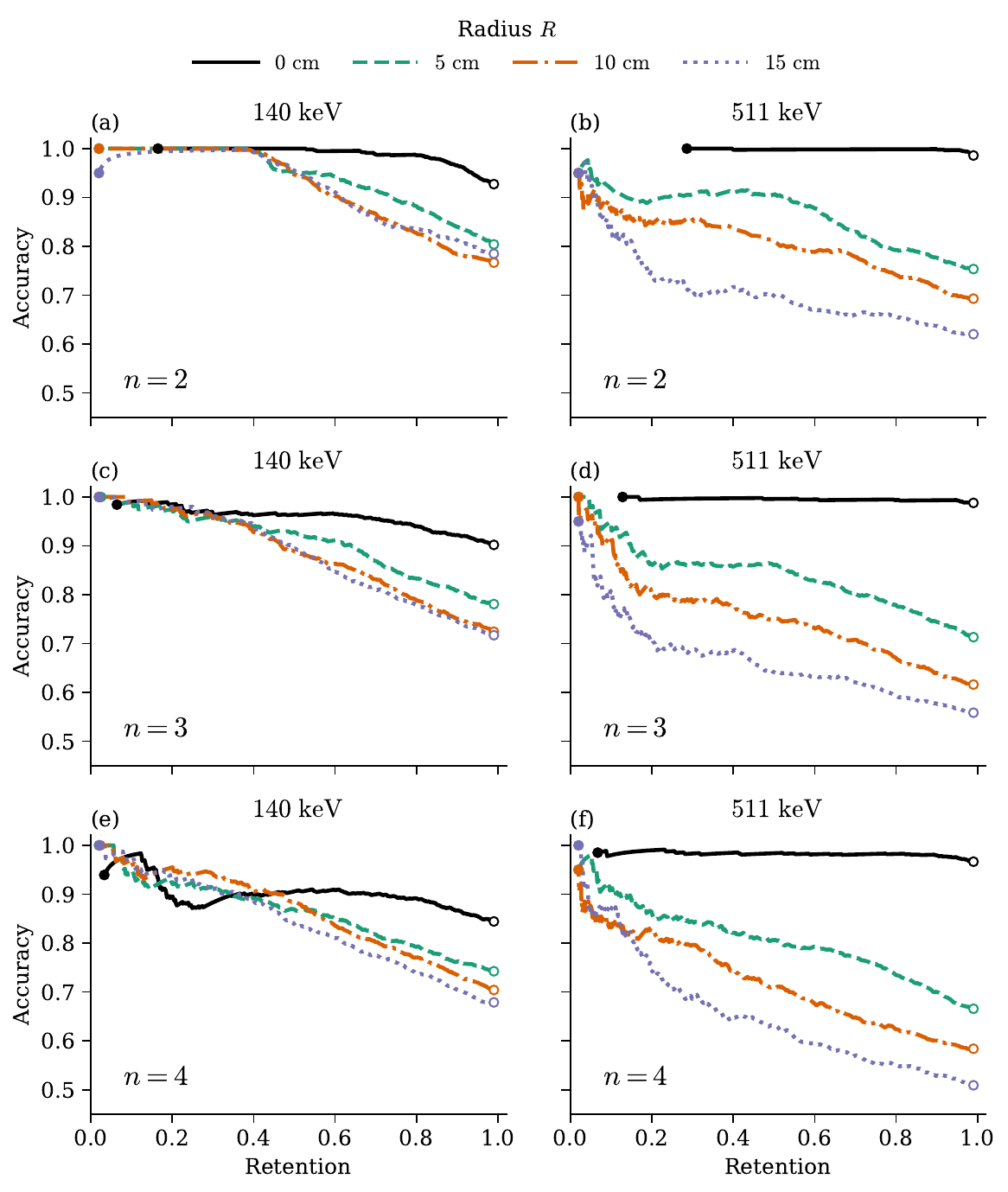}
\caption{Order-inference accuracy among retained events as a function of the retained event fraction under confidence filtering based on the MAP log-probability gap $\Delta \log p$. Rows correspond to interaction-chain lengths $n=2,3,4$, and columns correspond to source energies 140\,keV and 511\,keV. Curves represent different phantom radii. Open markers indicate the unfiltered case, while filled markers correspond to the strongest applied filtering.}
\label{fig:gap_threshold}
\end{figure}

Figure~\ref{fig:gap_threshold} shows the order-inference accuracy among retained events as a function of the retained event fraction under thresholding on the MAP log-probability gap $\Delta \log p$. For all configurations, increasing the threshold increases the accuracy among retained events while decreasing the retained fraction. This behavior is observed for both source energies and for all interaction-chain lengths considered.

\subsection{Scatter classification}

\begin{figure}[ht!]
\centering
\includegraphics[width=\linewidth]{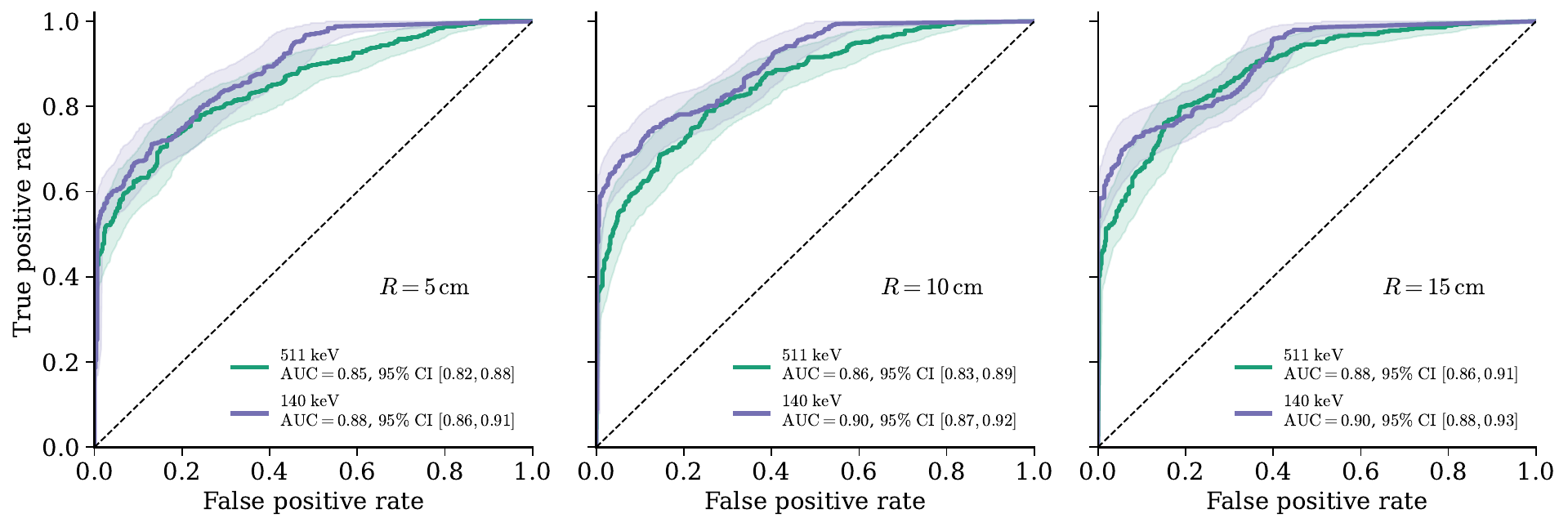}
\caption{ROC curves for scatter classification for phantom radii $R=5$, $10$, and $15$\,cm for both photon energies. The positive class is the primary class. Each ROC curve is computed from a balanced validation dataset containing 500 primary and 500 non-primary photons. Reported confidence intervals were obtained by bootstrap resampling over events within each evaluation dataset.}
\label{fig:roc_curves}
\end{figure}

\begin{figure}[ht!]
\centering
\includegraphics[width=0.6\linewidth]{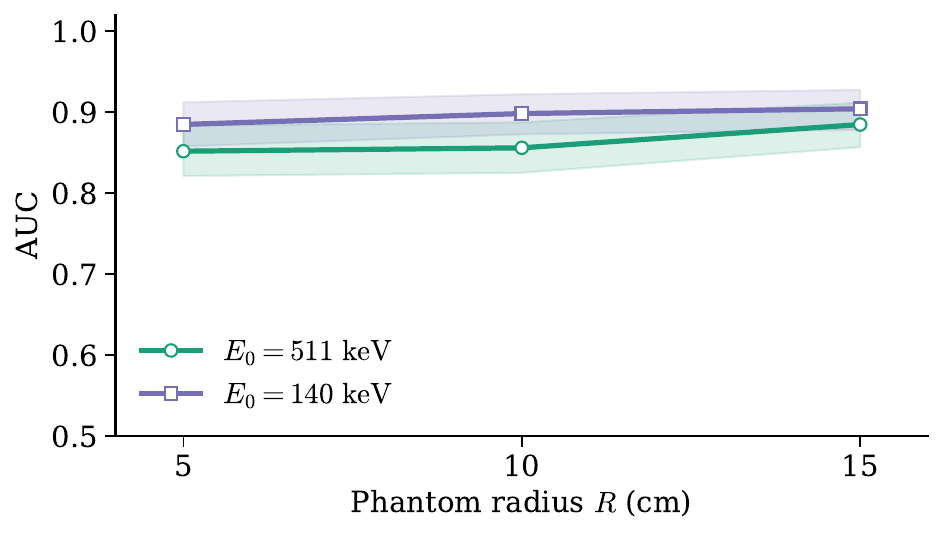}
\caption{Area under the ROC curve (AUC) for scatter classification as a function of phantom radius. The AUC values are computed from balanced validation datasets containing equal numbers of primary and non-primary photons.}
\label{fig:auc_vs_radius}
\end{figure}

Figure~\ref{fig:roc_curves} shows the receiver operating characteristic (ROC) curves for scatter classification for phantom radii $R=5$, $10$, and $15$\,cm for both source energies. In this evaluation, the positive class is the primary class. The ROC curves are computed on balanced validation datasets containing 500 primary and 500 non-primary photons for each configuration. The corresponding area under the curve (AUC) values are shown in Figure~\ref{fig:auc_vs_radius}. Across all configurations considered, the AUC exceeds 0.85.

\section{Discussion}
\label{sec:discussion}

The present results support the feasibility of probabilistic inference directly from the measured interaction sequence. Within the same framework, the recorded interactions were used both to infer causal interaction order and to distinguish between primary and non-primary photons. In this sense, the study serves as a proof of principle that interaction data recorded in a silicon detector contain enough information to support several inference tasks within one common probabilistic model.

For interaction order inference, the highest accuracies were obtained in the case without a phantom. For $511~\mathrm{keV}$, the accuracy in this case exceeded $95\%$ for all interaction-chain lengths considered. For both source energies, the accuracy decreased with increasing phantom radius and with increasing interaction-chain length. The dependence on phantom radius is consistent with the broader and more uncertain photon state at detector entry induced by object scatter. The dependence on interaction-chain length is also expected, since the number of admissible permutations increases as $n!$. Even in the most difficult case considered, $E_0 = 511~\mathrm{keV}$, $R = 15~\mathrm{cm}$, and $n = 4$, the ordering accuracy remained well above the random baseline of $1/4! \approx 4.2\%$, indicating that the likelihood model captures informative structure in the measured interaction sequence.

A difference between the two source energies was also observed. In the absence of a phantom, the $511~\mathrm{keV}$ datasets yielded higher ordering accuracy than the $140~\mathrm{keV}$ datasets. When object scatter was present, this trend reversed in the present simulations, with the $140~\mathrm{keV}$ datasets giving higher accuracies than the $511~\mathrm{keV}$ datasets. This suggests that the relative difficulty of the ordering problem depends both on source energy and on the additional uncertainty introduced by the phantom.

The posterior log-probability gap provided a useful confidence measure for interaction ordering. Across all datasets, thresholding on this quantity increased the ordering accuracy among retained events while reducing the retained fraction. Thus, the posterior itself provides a natural mechanism for selecting events with higher confidence. The relative gain obtained by thresholding was broadly similar across configurations, although some deviations from strict monotonicity were observed in a few cases. These are likely due to finite sample fluctuations and possibly also to limitations of the present likelihood model. The main result, however, is that the framework provides not only an ordering estimate, but also a confidence measure for each event.

For scatter classification, the likelihood-ratio test gave AUC values above $0.85$ for all phantom radii and both source energies. Since these results were obtained on class-balanced validation datasets, they should be interpreted primarily as showing separability between primary and non-primary events on the basis of the measured interaction sequence, rather than as direct estimates of classification performance in a realistic imaging setting where class proportions vary with object size and source energy. With that interpretation, the results indicate that the interaction sequence contains information that can be used to distinguish photons that entered the detector directly from photons that scattered in the object before detection.

In relation to previous work on event reconstruction and event selection in Compton systems, the present study uses a probabilistic formulation based on photon transport physics and detector response in a silicon-detector setting, and applies it to two specific inference tasks. The main point is therefore not only that interaction ordering and scatter discrimination can each be studied individually, but that both can be formulated within the same probabilistic description of the interaction chain.

The study is based on simulated data and an idealized detector model. In particular, the formulation assumes perfect position measurement and neglects effects such as finite pixel resolution, charge sharing, and electronic noise. Scatter classification was also evaluated under the assumption that the interaction order is known. These simplifications were useful for isolating the information content of the measured interaction sequence, but they should be relaxed in future work.

A natural next step is to extend the framework to additional latent variables and to more realistic detector-response assumptions. In particular, the same probabilistic formulation could in principle be used not only for interaction ordering and scatter discrimination, but also for inference of quantities such as the incident direction or the photon energy at detector entry. More broadly, the framework may provide a basis for event selection, event weighting, or other likelihood-based processing steps before image reconstruction.

Overall, the present results support probabilistic modeling of interaction chains as a proof of principle for inference from measured interaction data in silicon-based Compton imaging.
\section{Conclusion}

We have presented a probabilistic framework for modeling photon interaction chains in a silicon detector. The model combines photon transport physics with a detector response model and assigns likelihood values to candidate ordered interaction sequences on the basis of measured interaction positions and deposited energies.

Two inference tasks were studied. First, the framework was used to infer the causal interaction order of unordered detector events. Second, it was used to distinguish primary photons from photons that had scattered before entering the detector. In the simulated datasets considered here, the ordering accuracy was highest in the absence of a phantom and decreased with increasing phantom radius and interaction-chain length. The posterior log-probability gap provided a useful confidence measure for selecting events with more reliable ordering estimates. For scatter classification, the results showed clear separability between primary and non-primary events, with AUC values above $0.85$ for all phantom radii considered.

The present study should be viewed primarily as a proof of principle. Even under the simplified detector assumptions used here, the measured interaction sequence contains enough information to support probabilistic inference directly at the event level within one common framework. This suggests that probabilistic modeling of interaction chains may be a useful basis for more complete event processing in silicon-based Compton imaging.

Future work should extend the model to more realistic detector-response assumptions and to additional latent variables. In particular, the same framework could in principle be used for inference of quantities such as incident direction or photon energy at detector entry, and could serve as a basis for event selection, event weighting, or other likelihood-based processing steps before image reconstruction.
\appendix
\section{Explicit likelihood expression}

In this appendix, we give an explicit expression for the likelihood $p(m_\tau \mid E_0, \mathbf{u}_0')$ of an ordered measurement sequence under the transport and measurement models defined in the main text. The purpose is to show how the general marginal likelihood simplifies when the interaction positions are assumed to be measured exactly.

Under the assumption of perfect position measurement, the interaction positions are fixed by the data. For each interaction index $i=1,\dots,n$, the latent position coincides with the measured position, that is, $\mathbf{r}_i = \hat{\mathbf{r}}_{\tau(i)}$. For a given ordering $\tau$, the photon directions between consecutive interactions are therefore determined by the ordered geometry. In particular, the propagation direction between interactions $i-1$ and $i$ is given by $\mathbf{u}_{i-1} = (\mathbf{r}_i - \mathbf{r}_{i-1}) / \|\mathbf{r}_i - \mathbf{r}_{i-1}\|$, and the corresponding scattering angles satisfy $\cos\theta_i = \langle \mathbf{u}_{i-1}, \mathbf{u}_i \rangle$ for $i=1,\dots,n-1$.

Given the incident energy $E_0$, the post-interaction energies are then determined recursively by the Compton equation. For each interaction $i=1,\dots,n-1$, the outgoing energy is $E_i = \frac{mc^2 E_{i-1}}{mc^2 + E_{i-1}(1-\cos\theta_i)}$, and the deposited energy is $\varepsilon_i = E_{i-1} - E_i$. Thus, for fixed $(E_0,\mathbf{u}_0')$ and ordering $\tau$, the entire latent interaction chain is uniquely determined. We denote this deterministic chain by $q(m_\tau,q_0)$.

The likelihood can therefore be written as
\[
p(m_\tau \mid E_0,\mathbf{u}_0')
=
p(q(m_\tau,q_0)\mid q_0)\,
p(m_\tau \mid q(m_\tau,q_0)),
\]
where the first factor is the transport contribution and the second factor is the measurement contribution.

The transport contribution follows from the Markov structure of the photon propagation model:
\[
p(q \mid q_0)=\prod_{i=1}^{n} p(q_i \mid q_{i-1}).
\]
Evaluating this expression along the deterministic chain gives
\[
p(q(m_\tau,q_0)\mid q_0)
=
\prod_{i=1}^{n} p(q_i \mid q_{i-1}).
\]
For $i=1,\dots,n-1$, the transition density factorizes as
\[
p(q_i \mid q_{i-1})
=
p(E_i \mid E_{i-1})\,
p(\mathbf{r}_i \mid \mathbf{r}_{i-1},E_{i-1}),
\]
whereas the final transition is given by the last-interaction expression defined in the main text, which includes both Compton scattering and photoelectric absorption.

The measurement contribution is
\[
p(m_\tau \mid q)
=
\prod_{i=1}^{n} p(m_{\tau(i)} \mid q_{i-1},q_i).
\]
For interactions $i=1,\dots,n-1$, each factor is written as
\[
p(m_{\tau(i)} \mid q_{i-1},q_i)
=
p(\hat{\varepsilon}_{\tau(i)} \mid \varepsilon_i,\cos\theta_i)\,
\delta(\hat{\mathbf{r}}_{\tau(i)}-\mathbf{r}_i).
\]
For the final interaction, the measurement term is written as
\[
p(m_{\tau(n)} \mid q_{n-1},q_n)
=
p(\hat{\varepsilon}_{\tau(n)} \mid \varepsilon_n)\,
\delta(\hat{\mathbf{r}}_{\tau(n)}-\mathbf{r}_n).
\]
Since the positions are fixed by the measurements, the delta functions evaluate to unity along the deterministic chain. The measurement contribution therefore reduces to a product of energy likelihood terms.

Combining the transport and measurement contributions yields
\begin{equation}
\begin{aligned}
p(m_\tau \mid E_0,\mathbf{u}_0')
&=
\left[
\prod_{i=1}^{n} p(q_i \mid q_{i-1})
\right]
\left[
\prod_{i=1}^{n-1}
p(\hat{\varepsilon}_{\tau(i)} \mid \varepsilon_i,\cos\theta_i)
\right] \\
&\qquad \times
p(\hat{\varepsilon}_{\tau(n)} \mid \varepsilon_n),
\end{aligned}
\end{equation}
where all quantities are evaluated along the deterministic chain defined by $(E_0,\mathbf{u}_0',m_\tau)$.

\subsection{Algorithmic evaluation}

The explicit factorization above leads directly to a sequential evaluation of the log-likelihood for a fixed ordered event. Algorithm~\ref{alg:event_likelihood} summarizes this computation. The ordered interaction positions are first fixed by the data, after which the propagation directions, scattering angles, transport terms, and measurement terms are evaluated in sequence along the chain.

\begin{algorithm}[h]
\caption{Log-likelihood evaluation for a fixed ordered event}
\label{alg:event_likelihood}
\begin{algorithmic}[1]
\Require Ordered measurement chain $m_\tau$, incident energy $E_0$, incident direction $\mathbf{u}_0'$
\Ensure Log-likelihood $\ell$

\State Set $\mathbf{r}_i \gets \hat{\mathbf{r}}_{\tau(i)}$ for $i=1,\dots,n$
\State Determine $\mathbf{r}_0$ from $\mathbf{r}_1$, $\mathbf{u}_0'$, and detector geometry
\State Initialize $\ell \gets 0$

\For{$i=1,\dots,n$}
    \State $\mathbf{u}_{i-1} \gets (\mathbf{r}_i-\mathbf{r}_{i-1}) / \|\mathbf{r}_i-\mathbf{r}_{i-1}\|$
\EndFor

\For{$i=1,\dots,n-1$}
    \State $\cos\theta_i \gets \langle \mathbf{u}_{i-1},\mathbf{u}_i \rangle$
    \State $E_i \gets \frac{mc^2 E_{i-1}}{mc^2 + E_{i-1}(1-\cos\theta_i)}$
    \State $\varepsilon_i \gets E_{i-1}-E_i$
    \State $\ell \gets \ell + \log p(E_i\mid E_{i-1})$
    \State $\ell \gets \ell + \log p(\mathbf{r}_i\mid \mathbf{r}_{i-1},E_{i-1})$
    \State $\ell \gets \ell + \log p(\hat{\varepsilon}_{\tau(i)}\mid \varepsilon_i,\cos\theta_i)$
\EndFor

\State Determine the final deposited energy $\varepsilon_n$ from the final transition model
\State $\ell \gets \ell + \log p(q_n\mid q_{n-1})$
\State $\ell \gets \ell + \log p(\hat{\varepsilon}_{\tau(n)}\mid \varepsilon_n)$

\State \Return $\ell$
\end{algorithmic}
\end{algorithm}
\section* {Acknowledgment}
This work was financed by the Swedish Research Council (grant 2022-03688) and the European Union (ERC, Si3, grant agreement 101142294). 

Funded by the European Union. Views and opinions expressed are however those of the authors only and do not necessarily reflect those of the European Union or the European Research Council Executive Agency. Neither the European Union nor the granting authority can be held responsible for them.

\bibliography{report}   
\bibliographystyle{spiejour}   

\listoffigures
\listoftables
\end{spacing}
\end{document}